\documentclass[preprints,article,accept,moreauthors,pdftex]{Definitions/mdpi}
\setitemize{parsep=6pt,itemsep=0pt,leftmargin=*,labelsep=5.5mm}
\setenumerate{parsep=6pt,itemsep=0pt, leftmargin=*,labelsep=5.5mm}
\setlist[description]{itemsep=0mm}

\firstpage{1} 
\pubvolume{xx}
\issuenum{1}
\articlenumber{5}
\pubyear{2020}
\copyrightyear{2020}
\history{Received: 17 December 2019; Accepted: 4 January 2020; Published: date}
\updates{yes} 

\usepackage{amssymb}
\usepackage{dsfont}
\usepackage{soul}
\usepackage{stmaryrd}
\SetSymbolFont{stmry}{bold}{U}{stmry}{m}{n}

\usepackage{faktor}
\usepackage{physics}

\usepackage[labelformat=simple]{subcaption}

\DeclareCaptionLabelFormat{subcaptionlabel}{\normalfont(\textbf{#2}\normalfont)}
\newcommand{\bigmid}{\:\middle|\:}
\newcommand{\range}[2]{\left\llbracket #1, #2 \right\rrbracket}

\newcommand{\up}{\uparrow}
\newcommand{\down}{\downarrow}

\DeclareMathOperator{\Var}{Var}

\tikzstyle{vertex} = [circle, fill = white, draw = black]

\newcommand{\includetikzpicture}[2][]{\includegraphics[{#1}]{graphics/#2.pdf}}

\newcommand{\threetoone}{$3\mathrm{-to-}1$ }
\newcommand{\onetothree}{$1\mathrm{-to-}3$ }

\Title{Dynamical Triangulation Induced by Quantum Walk}

\Author{Quentin Aristote $^{1,2}$, Nathana\"el Eon $^{1,*}$ and Giuseppe Di Molfetta $^{1,*}$}

\AuthorNames{Quentin Aristote, Nathana\"el Eon and Giuseppe Di Molfetta}

\address{%
$^{1}$ \quad  Aix-Marseille Université, Université de Toulon, CNRS, LIS, Marseille, 13397, France; quentin.aristote@ens.fr\\
$^{2}$ \quad \'Ecole Normale Supérieure, PSL University, Paris, 75005, France}

\corres{Correspondence: nathanael.eon@lis-lab.fr (N.E.); giuseppe.dimolfetta@lis-lab.fr (G.D.M.)}
\abstract{We present the single-particle sector of a quantum cellular automaton, namely a quantum walk, on a simple dynamical triangulated $2-$manifold. The triangulation is changed through Pachner moves, induced by the walker density itself, allowing the surface to transform into any topologically equivalent one. This model extends the quantum walk over triangular grid, introduced in a previous work, by one of the authors, whose space-time limit recovers the Dirac equation in (2+1)-dimensions. Numerical simulations show that the number of triangles and the local curvature grow as \texorpdfstring{$t^\alpha e^{-\beta t^2}$}{t\^a e\^{-b t\^2} }, where \texorpdfstring{$\alpha$}{alpha} and \texorpdfstring{$\beta$}{beta} parametrize the way geometry changes upon the local density of the walker, and that, in the long run, flatness emerges. Finally, we also prove that the global behavior of the walker, remains the same under spacetime random fluctuations.}


\begin{document}

\section{Introduction}
Quantum cellular automata (QCA) are the quantum extension of cellular automata (CA): cells can now be in a quantum superposition of states and their evolution is unitary, homogeneous (translation-invariant) and causal, \textit{i.e} it only depends on a finite neighborhood of the cell. Just like CA are Turing complete, that is any algorithm can be simulated by a CA, QCA can simulate any quantum algorithm \cite{arrighi2012partitioned}. Quantum walks (QWs) are the single-particle sector of QCA \cite{costa2018quantum}. Their study is blossoming, for two parallel reasons.

On the one hand, QWs have enabled the discovery of whole series of novel quantum computing algorithms for future quantum computers \cite{BooleanEvalQW,ConductivityQW}, or allow a natural and intuitive expression of those algorithms, for instance the Grover search \cite{guillet2019grover}.

On the other hand, QWs have also enabled the discovery of a whole series of novel quantum simulation schemes for the near-future simulation devices, and allow more intuitive expressions of these schemes \cite{Bialynicki-Birula,MeyerQLGI}. Quantum simulations are what led Feynman to introduce quantum computing in the first place \cite{FeynmanQC}. Although universal quantum computers are not yet a reality, QW-based quantum simulation devices have already been designed \cite{WernerElectricQW,Sciarrino} : the walker propagates on the square grid in such a way that the continuum limit of its behavior converges towards the physics equation that is to be simulated \cite{di2013quantum, di2016quantum,di2020quantum}. Those schemes are both \emph{(i)} stable numerically, even for classical computers\textemdash therefore converging as long as they are consistent \cite{ArrighiDirac} ; \emph{(ii)} simple discrete models of the physical phenomena that conserve unitarity, homogeneity, causality and sometimes even Lorentz-covariance \cite{arrighi2014discrete,DArianoLorentz}\textemdash and thus provide playgrounds to discuss foundational questions in Physics \cite{LloydQG}. In a nutshell, QWs are becoming a new language to express quantum physical phenomena.

Even as QWs provide playgrounds to express a large class of quantum physical phenomena, discrete graph dynamics aim at providing a framework for the study of discrete transformation of spacetime. It is well known that any curved manifold can be approximated by some discrete graph, in particular equilateral triangles \cite{regge1961general, ambjorn1997geometry}, or simplices in higher dimensions, whose characteristic lengths, may ultimately be shrunk to zero to recover a continuum theory. One question that is still open is how the matter can propagate on such dynamical triangulations. Since QWs are ideal in simulating the propagation of fermions, here we will investigate with a simple example, how to couple the propagation of QWs to a particular class of discrete graph dynamics. Indeed, although QWs have already been extensively studied when the grid is fixed, and classical cellular automata on dynamical grids in one and two dimensions have been introduced recently by \cite{LoveOneD,MeyerLove,ArrighiCayley}, to the best of our knowledge QWs have never been studied in cases where we allow the grid to change.

Our goal is thus to develop and study a QW on a $2\mathrm{-dimensional}$ discrete surface, where the dynamics of the surface depends on the walker's evolution, and \textit{viceversa}, which is reminiscent of the basic mechanism of Einstein field equations, governing spacetime dynamics. More particularly, we would like to have two already elegant theories to work together. On the one hand, the family of QWs considered here is the one described in \cite{ArrighiTriang}, as while being very simple, it has the Dirac Equation as a continuous limit and can easily be extended to account for a curved metric \cite{ArrighiIrregTriang}. Similar results had already been obtained on a rectangular lattice and generalized to higher dimensions \cite{di2013quantum, di2016quantum,di2020quantum}. On the other hand, Pachner moves are a kind of transformations of the surface which changes the triangulation but not the topology. The dynamics of a lattice, subject to Pachner moves, has already been studied, \textit{e.g.}, in the context of lattice gas models \cite{MeyerLove} or complex networks \cite{da2018complex, ArrighiNCMA}. Although we focus on the interaction between the dynamical grid and one walker, we believe that results are already rich enough, i.e. likely to display the main features of the multi-walkers theory, namely QCA.
  
The result obtained is a QW evolving on a dynamical surface, whose dynamics is induced by the probability amplitude of the walker itself. QWs with effective non-linearities induced by the walker itself, inspired by \cite{kivshar1989dynamics,kerr1987quantum}, have already been introduced in \cite{shikano2014discrete,di2015nonlinear}, proving analytically the convergence to the non-linear Dirac equation. The QW we present here, can be considered the first in which non-linearity is not introduced as an effective term, but is the result of an interacting microscopic dynamics. Although not every property we would expected of a QW is respected, for instance reversibility is lost due to the chosen transformation of the grid, we believe the results are still worth sharing since they allow two elegant discrete theories to work hand in hand.
 
The paper is organized as follows. In section \ref{sec:recap}, we remind the reader of the definition of the quantum walk over the triangular grid, introduced in \cite{ArrighiTriang}, and the definition of Pachner moves. In section \ref{sec:newqw}, we extend this quantum walker to a triangular grid subject to Pachner moves. In section \ref{sec:equation} we write the global equation that rules the behavior of the walker and the grid. Then, in section \ref{sec:simulations}, we show numerical simulations for different families of discrete surfaces and finally discuss local and global properties of the surface and the walker independently. 
 
\begin{figure}[ht!]
  \centering
  \includegraphics[width=0.75\linewidth]{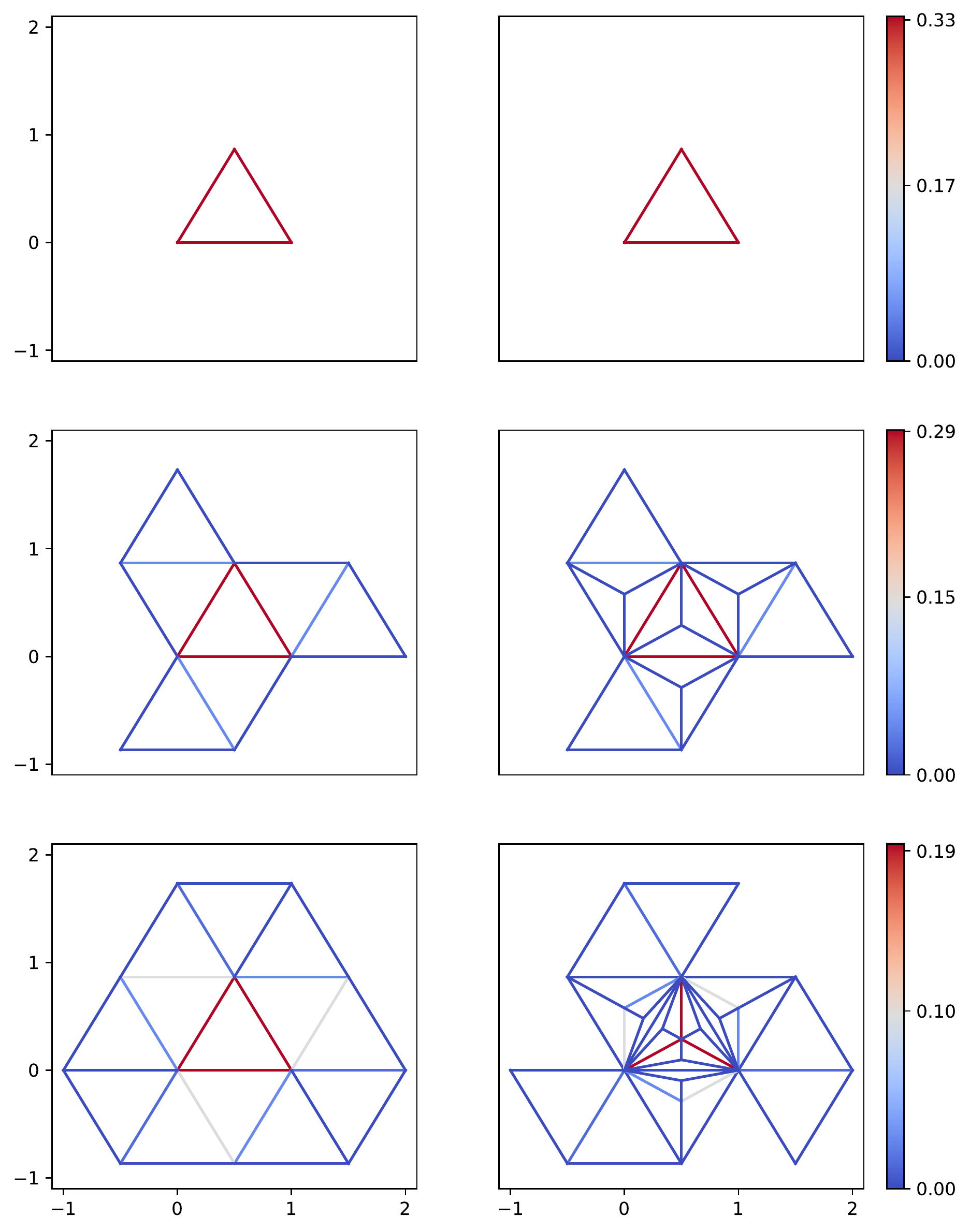}
  \caption{Time goes down with one time step between each figure vertically. $x$ and $y$ axis are spatial dimensions. On the left is the evolution of the usual quantum walk, on the right is the evolution of the walk with Pachner moves.}
  \label{fig:illustration}
\end{figure}

\section{Recap : Quantum Walk on the triangular grid and Pachner moves}
\label{sec:recap}

\subsection{Quantum walk on the triangular grid}

In order to define the quantum walk over the triangular grid, we consider the Hilbert space $\mathcal{H} = \mathcal{H}_e \otimes \mathcal{H}_s$ where $\mathcal{H}_e$ is spanned by the basis states $\ket{e}$ with $e$ an edge of the grid and $\mathcal{H}_s$ is spanned by the basis states $\ket{\up}$ and $\ket{\down}$. If $v$ is a triangle and $k \in \{1,2,3\} \cong \mathbb{Z}/3\mathbb{Z}$, we therefore write for a given time $t$:
\[ \psi(t,v,k) =
  \begin{pmatrix}
    \psi^\up(t,v,k) \\
    \psi^\down(t,v,k)
  \end{pmatrix}
\]
to represent the upper and lower components of the field, that we call \textit{spin}, carried by the $k\mathrm{-th}$ edge of $v$. You may notice that for two triangles sharing an edge, this edge contains twice the information (one for each triangle). This issue is addressed by associating each spin to a specific triangle. Let us label each triangle with the spin ($\up$ or $\down$) such that any two adjacent triangles have different spins. The most general step of time $\epsilon$ can then be divided into two substeps :
\begin{itemize}
\item first, rotate each triangle according to its label : if triangle $v$'s label is $s$, we set
  \[ \widetilde{\psi}^s\left(t + \frac{\epsilon}{2},v,k\right) = \widetilde{\psi}^s(t, v, k-1) \]
\item then, apply a unitary to each edge :
  \[\widetilde{\psi}(t + \epsilon, v, k) = W\widetilde{\psi}\left(t + \frac{\epsilon}{2}, v, k\right)\]
\end{itemize}
where $\widetilde{\psi}(t,v,k) = U_k\psi(t,v,k)$, and $U_k$ is an arbitrary element of U(2).

Since adjacent triangles have different labels, this can be rewritten as a whole as
\[ \widetilde{\psi}(t + \epsilon, v, k) = W
  \begin{pmatrix}
    \widetilde{\psi}^\up(t, v, k - 1) \\
    \widetilde{\psi}^\down(t, v, k - 1)
  \end{pmatrix}
\]

When $W$ and the $U_k$ are well-chosen, making three steps yields an equation that converges, when $\epsilon$ converges to $0$, to the $(2 + 1)\mathrm{-dimensional}$ Dirac Equation \cite{ArrighiTriang}, \textit{i.e.} the equation governing the dynamics of 1/2-spin fermions.

\subsection{Pachner moves}

\begin{figure}[!ht]
  \centering
  \makebox[0pt]{
    \includetikzpicture[width=.8\linewidth]{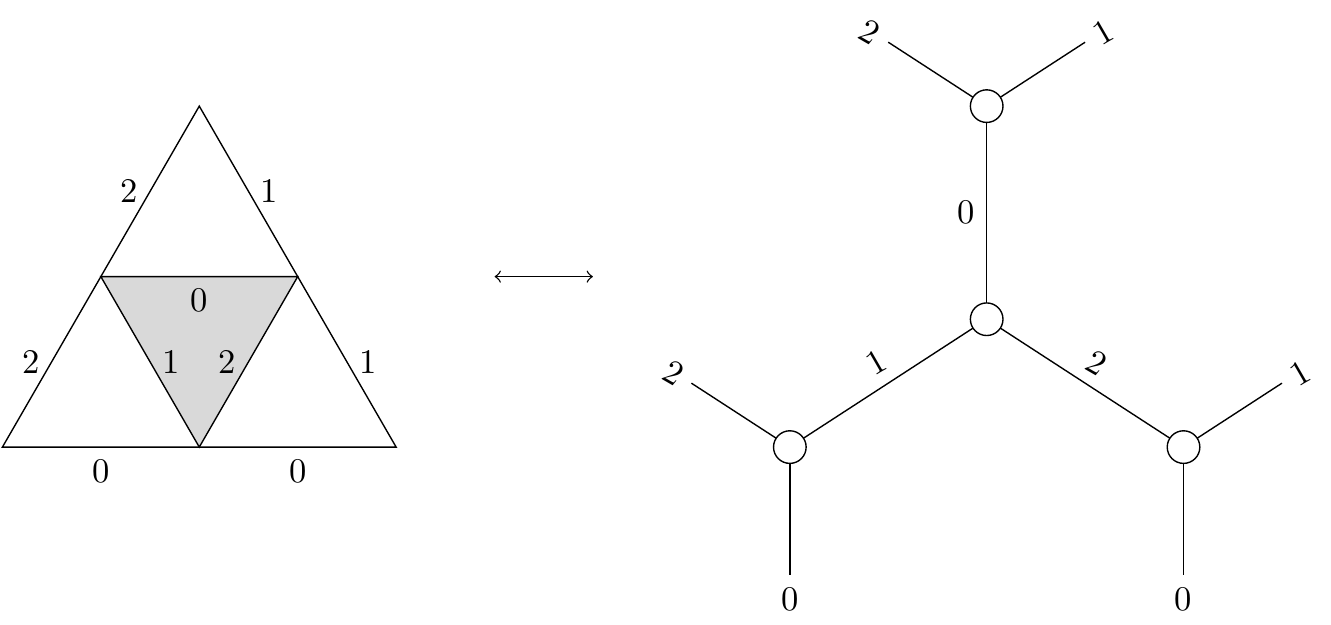}
  }
  \caption{Discrete manifold / graph duality}
  \label{fig:discrete_manifold_graph_duality}
\end{figure}

\begin{figure}[!ht]
  \centering
  \makebox[.29\textwidth] {
    \begin{subfigure}{.2\textwidth}
      \includetikzpicture[width=\linewidth]{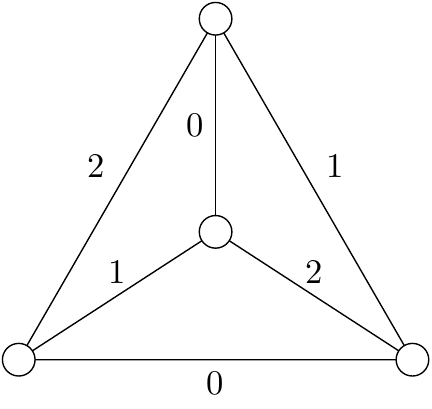}
      \caption{$\partial\Delta_3$}
      \label{fig:discrete_3_dim_sphere}
    \end{subfigure}
  }
  \makebox[.69\textwidth] {
    \begin{subfigure}{.6\textwidth}
      \includetikzpicture[width=\linewidth]{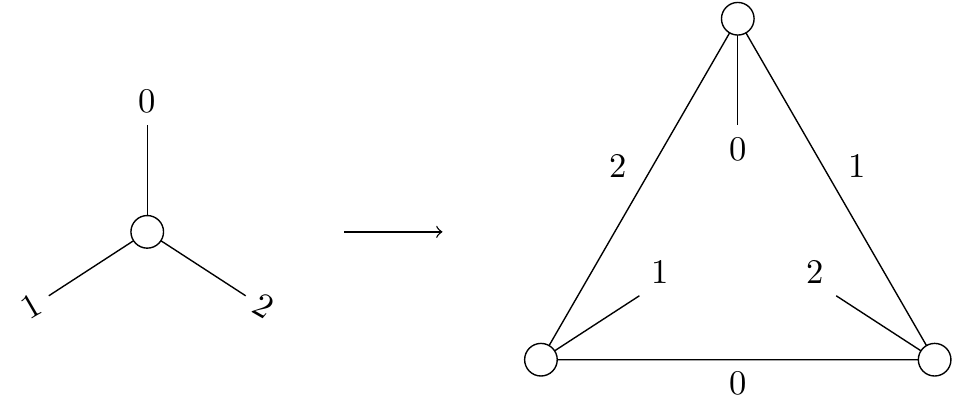}
      \caption{$1\mathrm{-to-}3$ Pachner move}
      \label{fig:1-to-3_pachner_move}
    \end{subfigure}
  }
  \caption{Pachner moves}
  \label{fig:pachner_moves}
\end{figure}

An $n\mathrm{-to-}m$ Pachner move is done by taking a subset of $n$ triangles of a manifold and replacing it by its complementary in the discrete $(n+m)\mathrm{-dimensional}$ sphere, $\partial\Delta_{n + m}$. It is easier to visualize them in the dual setting of graphs (each triangle of the complex is a vertex and each side of the triangle is an edge between the two triangles it separates, see Fig.\ref{fig:discrete_manifold_graph_duality}) \cite{ArrighiNCMA}, as in Fig.\ref{fig:pachner_moves}.

Pachner moves are remarkable because any discrete manifold homeomorphism can be seen as a finite sequence of Pachner moves, therefore they do not change the topology of the surface. Thus by allowing Pachner moves on our grid we actually allow it to transform into any topologically equivalent discrete manifold. 

Since $\partial\Delta_3$ is a tetrahedron, one can see $1\mathrm{-to-}3$ Pachner moves as creating a well in the grid and $3\mathrm{-to-}1$ Pachner moves as removing those wells.

\section{Coupling the Quantum Walker with Pachner moves}\label{sec:newqw}

\subsection{The grid}

In section \ref{sec:recap}, the triangles of the grid were labeled with a spin ($\up$ or $\down$), telling us which component of the field the triangle carries. An important property is that two adjacent triangles cannot be labeled with the same spin, as both components of the walker on their shared edge must be propagated. It does not hold true when taking Pachner moves into account as $1\mathrm{-to-}3$ Pachner moves create $3\mathrm{-cycles}$ (it is impossible to find a $2\mathrm{-coloration}$ of a graph with $3\mathrm{-cycles}$). We thus introduce the following set of labels :
\[ \Sigma = \{ (\up, \up, \up), (\down, \down, \down), (\up, \up, \down), (\down, \down, \up) \} \]
A triangle being labeled $(s_1,s_2,s_3)$ means that it carries the $s_k$ component of its $k\mathrm{-th}$ side for each $i \in \mathbb{Z}/3\mathbb{Z}$. The triangular grid can thus be labeled the same way as in section \ref{sec:recap}, by identifying $\up$ with $(\up, \up, \up)$ and $\down$ with $(\down, \down, \down)$.

We therefore define our new grid as a simplicial complex where each triangle is labeled with an element of $\Sigma$, or, similarly, as a labeled graph  where each vertex represents a triangle, each edge represents a side, the vertices are labeled with elements of $\Sigma$ and the edges are labeled with elements of $\pi = \mathbb{Z}/3\mathbb{Z}$.

\subsection{Evolution of the grid under Pachner moves}

\begin{figure}[ht!]
  \centering
  \includetikzpicture{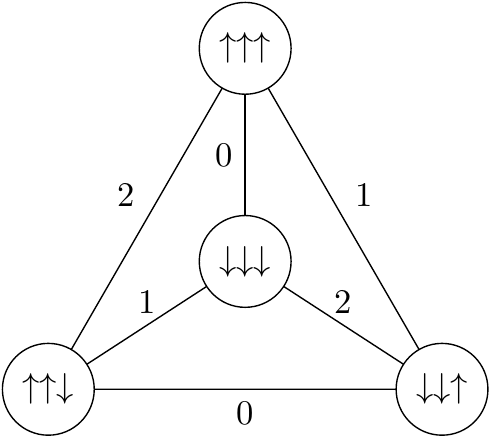}
  \caption{$\partial\Delta_3$ labeled}
  \label{fig:labeled_3_dim_sphere}
\end{figure}

In the simplicial complex setting, Pachner moves can be seen as replacing a subset of the complex with its complementary in a simplicial sphere $\partial\Delta_n$. To define them in our labeled simplicial complex setting, we label the $3\mathrm{-dimensional}$ sphere as in Fig.\ref{fig:labeled_3_dim_sphere}.

\begin{figure}[ht!]
  \centering
  \includetikzpicture[width=\linewidth]{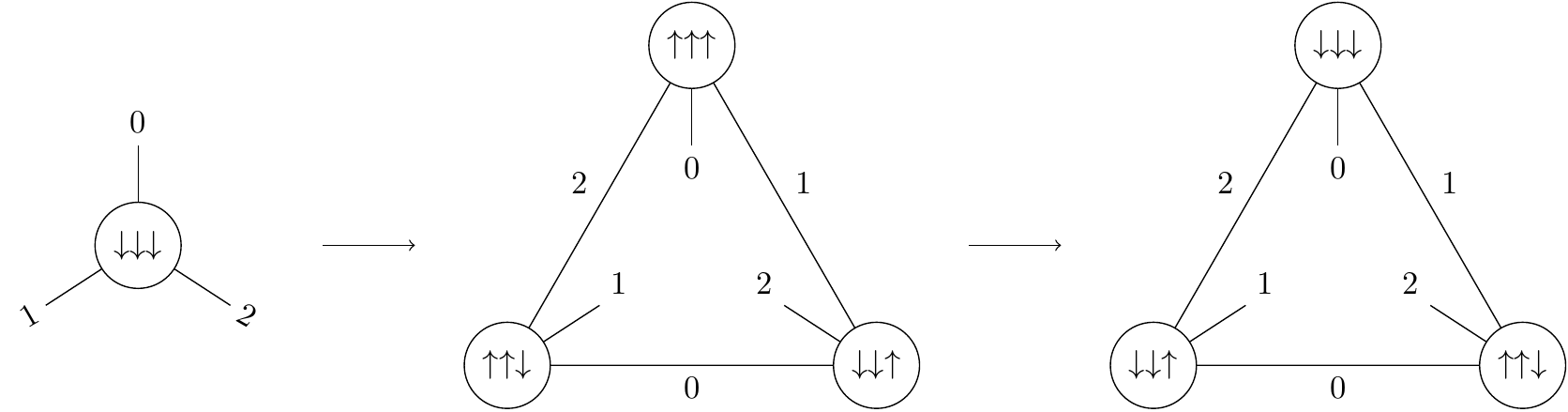}
  \caption{1-to-3 Pachner move in the dual setting of graphs}
  \label{fig:labeled_1-to-3_pachner_move}
\end{figure}

Making a Pachner move now amounts to taking the complementary in the labeled version of $\partial\Delta_3$ and then inverting the labels ($\up$ becomes $\down$ and $\down$ becomes $\up$) so that two adjacent triangles each carry a different component of the field along their shared edge (Fig.\ref{fig:labeled_1-to-3_pachner_move}).

Notice that this means the $1\mathrm{-to-}3$ Pachner move is still the inverse of the $3\mathrm{-to-}1$ Pachner move and that the $2\mathrm{-to-}2$ Pachner move is still its own inverse.

We now show that we can make Pachner moves whenever we want to. 

\begin{Lemma}
  Two adjacent triangles always have different labels.
\end{Lemma}

\begin{proof}
  Consider two adjacent triangles with labels $(s_1, s_2, s_3)$ and $(s_1', s_2', s_3')$, sharing their $k\mathrm{-th}$ side. We made sure that after any sequence of Pachner moves two adjacent triangles each carry a different component along their shared edge. Therefore $s_k \neq s_k'$, hence $(s_1, s_2, s_3) \neq (s_1',s_2',s_3')$ : the two triangles have different labels.
\end{proof}

\begin{Corollary}
  It is always possible to apply a Pachner move on a complete subgraph of the grid, as long as it has less than $4$ vertices.
\end{Corollary}

\subsection{The quantum walker}

At each timestep $t$, the field $\widetilde{\psi}$ and the grid evolve in the following order :
\begin{itemize}
\item first, each triangle rotates its internal components : if triangle $v$'s label is $(s_1,s_2,s_3)$,
  \[ \widetilde{\psi}^{s_k}\left(t + \frac{\epsilon}{2}, v, k\right) = \widetilde{\psi}^{s_{k-1}}(t, v, k - 1) \]
\item second, we apply $W$ to each edge :
  \[ \widetilde{\psi}(t + \epsilon, v, k) = W\widetilde{\psi}\left(t+\frac{\epsilon}{2}, v, k\right) \]
\end{itemize}

Notice that, at this point, if the grid is the standard triangular grid, the walker evolves exactly as in section \ref{sec:recap}, as it is illustrated in Fig. \ref{fig:illustration}. 

\begin{itemize}
\item third, we apply the Pachner moves for this timestep.  
\end{itemize}

\subsection{Evolution of the walker during \texorpdfstring{$1\mathrm{-to-}3$}{1--to--3} and \texorpdfstring{$2\mathrm{-to-}2$}{2--to--2} Pachner moves}

\begin{figure}[ht!]
  \centering
  \includetikzpicture[width=\linewidth]{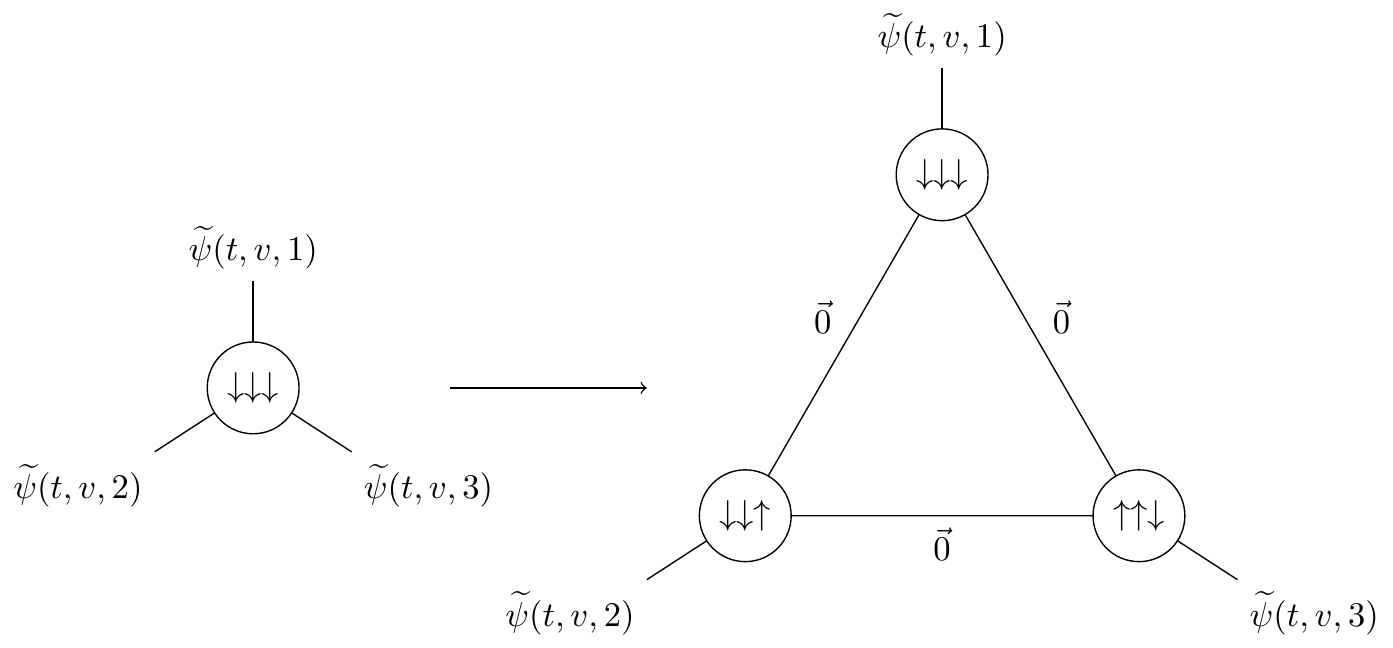}
  \caption{Evolution of the walker during a $1\mathrm{-to-}3$ Pachner move}
  \label{fig:walker_1-to-3}
\end{figure}

A $1\mathrm{-to-}3$ Pachner move can be seen as adding three new edges inside a triangle. We choose here to make  $\widetilde{\psi}$ stay the same on the edges that already existed before the Pachner move, and to set it to $0$ on the newly created edges (so that we still have $\norm{\psi}^2 = 1$) (Fig.\ref{fig:walker_1-to-3}). This is one choice among many, for instance one may have distributed the value of $\widetilde{\psi}$ between all edges. The choice made here has the advantage to be very simple.

\begin{figure}[ht!]
  \centering
  \includetikzpicture[width=\linewidth]{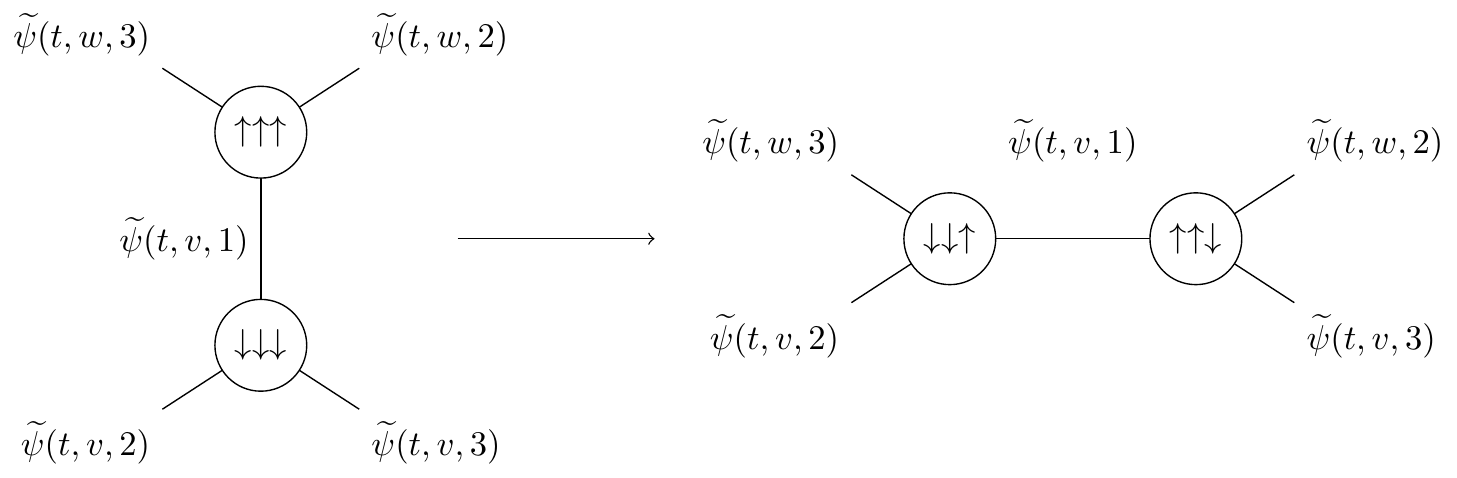}
  \caption{Evolution of the walker during a $2\mathrm{-to-}2$ Pachner move}
  \label{fig:walker_2-to-2}
\end{figure}

Similarly, a $2\mathrm{-to-}2$ Pachner move can be seen as changing which edges are part of the same triangle, without changing the edges themselves, as illustrated in Fig. \ref{fig:walker_2-to-2}. It therefore makes sense to have $\widetilde{\psi}$ stay the same on those edges.

\subsection{Evolution of the walker during \texorpdfstring{$3\mathrm{-to-}1$}{3--to--1} Pachner moves}

The evolution during $3\mathrm{-to-}1$ Pachner moves is more difficult to define. Indeed, such a move deletes edges and thus may break reversibility through a loss of information. Multiple choices of evolution for the walker are possible, for instance the values of $\widetilde{\psi}$ on the edges which will be deleted could be dispatched among the outer edges of the triangle containing them.

The constraint of not deleting any value of $\widetilde{\psi}$ implies that the evolution has to take into account the value of $\widetilde{\psi}$ on an infinite number of edges. Indeed, if $\widetilde{\psi}$ changes only on a finite number of edges we come back to the situation where some values will be mixed together. 
This is a problem as it would be best that Pachner moves only act locally, that is on a finite number of edges.

\begin{figure}[ht!]
  \centering
  \makebox[0pt]{
    \includetikzpicture{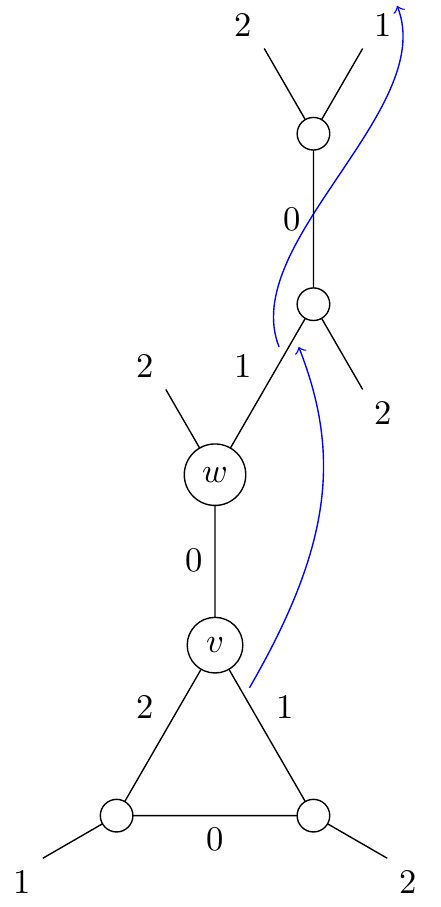}
  }
  \caption{Translation of the component internal to $w$ on its first side during a $3\mathrm{-to-}1$ Pachner move}
  \label{fig:walker_3-to-1}
\end{figure}

 To solve this problem we make the assumption that at time $t = 0$ (and therefore at any time $t$) the walker is in a superposition of a finite number of states, that is $\widetilde{\psi}(t = 0)$ is equal to $0$ on a cofinite set of edges. This way, it is possible to have an operator that acts on a vector space of infinite dimension while having it change only a finite number of values in practice.
This can for example be done with a translation. The most intuitive one we found is depicted on Fig.\ref{fig:walker_3-to-1}. If three triangles make a $3\mathrm{-cycle}$ onto which a $3\mathrm{-to-}1$ Pachner move is applied, then any internal component carried by one of those triangles (say, $v$) on one of its edges internal to the $3\mathrm{-cycle}$ (say, its $k\mathrm{-th}$ side) is translated along the side of $v$ that goes out of the $3\mathrm{-cycle}$ into the neighboring triangle $w$, and then is translated again along the $k\mathrm{-th}$ side of $w$ to finally replace the internal component of the neighbor of $w$ on its $k\mathrm{-th}$ side. The old value of this component is then translated along the same edges and replaces another internal component that also gets translated and replaces another value, and so forth.

Once the translation is done, the edges internal to the $3\mathrm{-cycle}$ can be deleted as the information they carried was sent outside of the $3\mathrm{-cycle}$.

It does make sense physically to have a triangle influence another one which is two edges away in the graph as the geometrical distance between the triangles is still one (the triangles share a vertex).
Notice that such an evolution can only be defined on an infinite grid. Otherwise, the evolution of $\widetilde{\psi}$ for a \threetoone Pachner move would be an infinite loop.

\begin{Lemma}
  Consider the infinite triangular grid and apply a finite number of $1\mathrm{-to-}3$ and $3\mathrm{-to-}1$ Pachner moves. Then any triangle of the grid appears at most once in the sequence of the triangles visited during a translation along two edges, as previously described.
\end{Lemma}

\begin{figure}[ht!]
  \centering
  \includetikzpicture{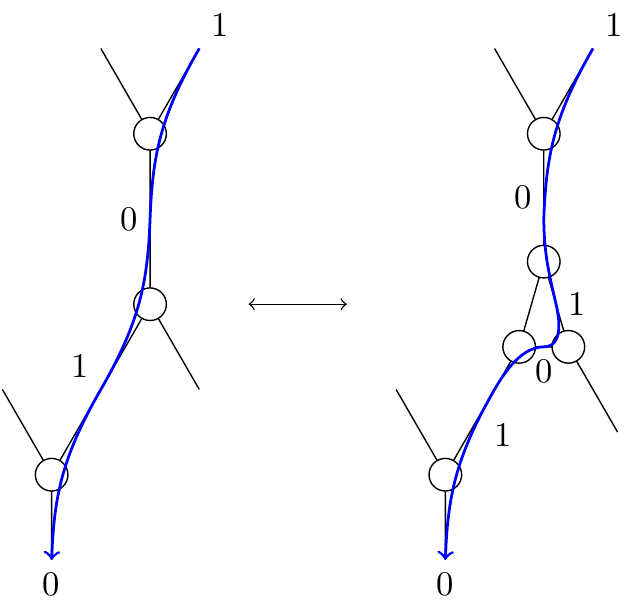}
  \caption{$1\mathrm{-to-}3$ and $3\mathrm{-to-}1$ Pachner have no influence on cycles}
  \label{fig:pachner_moves_and_cycles}
\end{figure}

\begin{proof}
  It is true for the infinite triangular grid : a translation along two edges amounts to translating along a fixed vector, therefore any triangle is visited at most once. Let us now consider a grid where this holds true. Then applying a $1\mathrm{-to-}3$ or a $3\mathrm{-to-}1$ Pachner move keeps this property true. Indeed, applying such a Pachner move does not destroy any cycle, as can be seen in Fig.\ref{fig:pachner_moves_and_cycles}. If we were to find a cycle in the new grid, we could apply the inverse Pachner move and get the original grid, without having destroyed the cycle. Therefore the cycle already existed in the original grid, which contradicts our assumption.

  By induction, the property thus holds true after a finite number of such Pachner moves.
\end{proof}

Unfortunately this does not hold true anymore if we consider $2\mathrm{-to-}2$ Pachner moves, as they introduce cycles. We were not able to find any evolution for $3\mathrm{-to-}1$ Pachner moves that worked when considering $2\mathrm{-to-}2$ Pachner moves as well. We therefore decided to restrict ourselves to $1\mathrm{-to-}3$ and $3\mathrm{-to-}1$ Pachner moves in the following. 
However, if another evolution of the walker during a \threetoone Pachner moves is used, it may be possible to make it work when considering $2\mathrm{-to-}2$ Pachner moves.

At this point, it is important to note that locality, causality and reversibility are lost. Loss of locality comes from the fact that the evolution on a \threetoone Pachner move changes values of $\widetilde{\psi}$ arbitrarily far away from the edges that will be deleted. Such loss of locality implies a loss of causality since the impact of the \threetoone move will have an impact arbitrarily far away in a single time step. Finally, the evolution was expected to keep reversibility since it keeps every value of $\widetilde{\psi}$, however, it does not keep track of where the \threetoone Pachner move has been done and the evolution cannot be reversed.
However, in a more general point of view, after a finite time, the walk will not propagates faster than the speed of light since at some point no \threetoone move will be done anymore. One can check so by computing the moments of the distributions and see that it is never hyperbalistic in the intermediate and long run, although it may be hyperbalistic for the short-run. Even though reversibility is lost in the process, we choose this evolution because it keeps every value of $\widetilde{\psi}$ and preserves the norm. 

\subsection{When to make Pachner moves\label{sec:when}}

In practice, after a $1\mathrm{-to-}3$ or a $3\mathrm{-to-}1$ Pachner move, the walker has to go through more, respectively less, triangles to travel a fixed distance. A greater or, respectively, lesser refinement of the triangular surface can then be seen as a local contraction or dilatation of the discrete metric on which the walker propagates. A greater/minor slowdown of the walker in a given region of space, can therefore be associated with a greater/minor curvature of the spatial metric. At this point, we have adopted a typical general relativity principle, for which matter tells space how to curve, and curved space tells matter how to move. We adopted this point of view here, choosing to make a $1\mathrm{-to-}3$ Pachner move (creating a well) on a triangle $v$ whenever the probability to be on that triangle is above a threshold $\alpha$ (if $v$ is labelled $(s_1,s_2,s_3)$, whenever $\sum_{k = 1}^3\abs{\psi^{s_k}(t, v, k)}^2 > \alpha$), and to make a $3\mathrm{-to-}1$ Pachner move (removing a well) whenever the probability to be inside that well is below a threshold $\beta$ (if $u$ and $v$ are glued along their first side, $v$ and $w$ along their second side and $w$ and $u$ along their third side, we make a Pachner move whenever $\norm{\psi(t,u,1)}^2 + \norm{\psi(t,v,2)}^2 + \norm{\psi(t,w,3)}^2 < \beta$). 
The expected evolution of the geometry can be inferred from the values of $\alpha$ and $\beta$. For low $\alpha$, there will be a great number of wells, and the structure can thus be seen as very stretchable and easily deformable, while for high $\alpha$, the number of wells created will be small, and the structure can thus be seen as rigid. On the other hand $\beta$ corresponds to the speed at which wells will be removed: for a high $\beta$ a well will be removed the step right after its creation while for a low $\beta$ the well will only be removed when the density inside it is sufficiently low (i.e. when enough time has passed).  

\section{Discrete equation of the walker}

\label{sec:equation}

We now write the discrete equation of the walker, that is $\psi(t + \epsilon)$ as a function of $\psi(t)$. To do so, we first change the way we write $\psi$, as thinking of it as a function of the triangles is cumbersome to write the equations for the Pachner moves. Instead, we choose a triangle of the triangular grid to be its origin. We can keep track of it when it is subject to Pachner moves as a $1\mathrm{-to-}3$ Pachner moves always create a triangle with the same label as the triangle to which it is applied (this would not be true had we kept $2\mathrm{-to-}2$ Pachner moves) : this new triangle becomes the origin. This makes the dual graph a pointed graph, and therefore every triangle can be thought of as the language of words that correspond to a path from the origin to the triangle \cite{ArrighiCayley}.
\begin{Notation}
  For $v$ the language associated to a triangle, we denote its neighbor on the first side by
  $$v.1  = \{u.1 \mid u \in v\}$$
  where the $.$ operation is the concatenation of words.
\end{Notation}

We identify any $u \in v$ with $v$ itself.
\begin{Notation}
We introduce two ways of writing $\psi$:
\begin{itemize}
\item if triangle $v$'s label is $(s_1,s_2,s_3)$, we write:
    \[ \psi(t,v:k) = \psi^{s_k}(t,v,k) \]
    \[ \psi(t,v:k:v.k) = \psi(t,v,k) \]
    
\item if triangle $v$ carries the $\up$ component of its $k\mathrm{-th}$ side and its neighbor $v.k$ on that side carries the $\down$ component, we write:
  \[ \psi\left(t, \begin{array}{c} v \\ \hline k \\ \hline v.k \end{array} \right) =
    \begin{pmatrix}
      \psi(t,v:k) \\
      \psi(t,v.k:k)
    \end{pmatrix}
  \]
\end{itemize}
\end{Notation}
After rotating the triangles and applying $W$, we have:
\[\widetilde{\psi}\left(t + \epsilon, \begin{array}{c} v \\ \hline k \\ \hline w \\ \end{array}\right) = W \begin{pmatrix} \widetilde{\psi}(t, v:k - 1) \\ \widetilde{\psi}(t, w:k - 1)\end{pmatrix} = (WR\widetilde{\psi}(t))\left(\begin{array}{c} v \\ \hline k \\ \hline w \\ \end{array}\right)\]
(we see $W$ as both a $2\mathrm{-dimensional}$ operator when it acts on a single edge and an infinite dimensional operator when it acts on all the edges).

After applying all the $1\mathrm{-to-}3$ Pachner moves, we have:
\begin{align*}
  \widetilde{\psi}&\left(t + 3\epsilon, \begin{array}{c} \Pi_{i = 1}^nk_i(k_{i+1}k_i)^{\mathds{1}_{1 \rightarrow_{t + 2\epsilon} 3}\left(\Pi_{j = 1}^i k_j\right)} \\ \hline k \\ \hline \left(\Pi_{i = 1}^nk_i(k_{i+1}k_i)^{\mathds{1}_{1 \rightarrow_{t + 2\epsilon} 3}\left(\Pi_{j = 1}^i k_j\right)}\right)k \end{array} \right) = \\
                 &\left(1 - \mathds{1}_{1 \rightarrow_{t + 2\epsilon} 3}\left(\Pi_{i = 1}^n k_i\right)(1 - \delta_{k_n = k})\right)\widetilde{\psi}\left(t + 2\epsilon,\begin{array}{c} \Pi_{i = 1}^n k_i \\ \hline k \\ \hline \left(\Pi_{i = 1}^n k_i\right)k \end{array}\right)
  \end{align*}
This describes both how triangles are renamed (we add letters in the middle of the word for each Pachner move done on the corresponding path) and what becomes of $\psi$ (it becomes zero only if the triangle is subject to a Pachner move and if the edge referred to is now part of the new $3\mathrm{-cycle}$, otherwise it stays the same). The function $\mathds{1}_{1\rightarrow_t3}$ tells us whether a $1\mathrm{-to-}3$ Pachner move is to be applied on the triangle at time $t$ :
\[
  \mathds{1}_{1 \rightarrow_t 3}(v) = \left\{\begin{matrix}
      1 & \mathrm{if} & \sum_{k = 1}^3 \abs{\psi(t, v:k)}^2 > \alpha \\
      0 & \mathrm{otherwise} &
    \end{matrix}\right.
\]

Finally, let $3\mathrm{-cycles}(t)$ be the set
\[\left\{c = \{u:0, v:1, w:2, u:1:w, u:2:v, v:0:w\} \bigmid \sum_{\substack{s \in \{u,v,w\} \\ k \in \pi \\ s:k \notin c}} \abs{\psi(t + \epsilon, s:k)}^2 < \beta \right\}\]
Then, after the $3\mathrm{-to-}1$ Pachner moves,
\[\widetilde{\psi}(t + 2\epsilon) = \left(\Pi_{c \in 3\mathrm{-cycles}(t + \epsilon)}D_c \circ T_c\right)\widetilde{\psi}(t + \epsilon)\]
with
\begin{align*}
  (T_c \widetilde{\psi})(u:k) &= \delta_{\exists n \in \mathbb{N}^*, u.(k(k+1))^n : k+1 \in c}\widetilde{\psi}(u.k(k+1) :k) \\
               &+ \delta_{\exists n \in \mathbb{N}^*, u.(k(k+2))^n : k+2 \in c}\widetilde{\psi}(u.k(k+2) :k) \\
               &+ \delta_{\forall n \in \mathbb{N}, l \in \{k+1,k+2\}, u.(kl)^n : l \notin c}\widetilde{\psi}(u :k)
\end{align*}
and
\[ (D_c \widetilde{\psi})\left(u = \Pi_{i = 1}^n k_i:k\right) = \widetilde{\psi}\left(\Pi_{i \in I_c^u}k_i\right)\]
where
\begin{equation*}
  c^u = \left\{ i \in \range{1}{n} \bigmid \Pi_{j = 1}^i k_j : k_{i+1} : \left(\Pi_{j = 1}^i k_j\right)k_{i+1} \notin c\right\}
\end{equation*}
$T_c$ is the operator that translates the values of $\psi$, this is the operator which breaks locality and reversibility.  $D_c$ is the one that renames the triangles after deleting the edges. Notice that we did not specify in which order the $3\mathrm{-cycles}$ $c$ should be taken : we choose to do so in descending probability, which breaks the rotation symmetry but does not influence the macroscopic behavior of the walk (as can be seen through the simulations). 

Putting all those equations together gives
\begin{align}
  \widetilde{\psi}&\left(t + 3\epsilon, \begin{array}{c}
                   \Pi_{i = 1}^nk_i(k_{i+1}k_i)^{\mathds{1}_{1 \rightarrow_t 3}\left(\Pi_{j = 1}^i k_j\right)} \nonumber \\
                   \hline k \\
                   \hline \left(\Pi_{i = 1}^nk_i(k_{i+1}k_i)^{\mathds{1}_{1 \rightarrow_t 3}\left(\Pi_{j = 1}^i k_j\right)}\right)k
                              \end{array} \right) = \\
  &\left(1 - \mathds{1}_{1 \rightarrow_t 3}\left(\Pi_{i = 1}^n k_i\right)(1 - \delta_{k_n = k})\right) \times
  \left(\left(\Pi_{c \in 3\mathrm{-cycles}}D_c \circ T_c\right)WR\widetilde{\psi}(t)\right)\left(\begin{array}{c}
                                                                               \Pi_{i = 1}^n k_i \\
                                                                               \hline k \\
                                                                               \hline \left(\Pi_{i = 1}^n k_i\right)k
                                                                             \end{array}\right)
                                                                             \label{eq: EQFIN}
\end{align}
Notice that we choose to apply the $3\mathrm{-to-}1$ Pachner moves before the $1\mathrm{-to-}3$ ones : this is because doing the converse would mean instantly deleting the newly created $3\mathrm{-cycles}$.

The equations are non-linear and not analytically soluble. Still, we can find some limite cases, for instance that choosing $\alpha = 1$ yields the quantum walk over the triangular grid from section \ref{sec:recap} (since $3\mathrm{-cycles}(t)$ is always empty) or that choosing $\beta = 0$ amounts to overlooking $3\mathrm{-to-}1$ Pachner moves. An extreme condition is for $\beta = 1$ and $\alpha=0$: the QW will be trapped in an ever deepening well. Notice that, as the probability is preserved and therefore finite, the number of PMs that could be carried out in a given region of space would remain finite in general. However for $\beta = 1$ and $\alpha=0$, the grid will be refined an arbitrary large number of times to create a hole of very deep potential well and, in principle we could have an arbitrary large number of refining in a finite region of space, which will certainly violate the principle of finite density of information \cite{arrighi2013principle}. To prevent such a situation, without imposing any cut-off on the grid, e.g., a Planck length scale, we choose $\beta$ to be proportional to $\alpha$, choice which excludes the above extreme condition. This is also reasonable since, making an analogy with mechanics, usually a surface has a single elasticity constant that is involved in both stretching and relaxation. Thus, $\alpha$ will be the only parameter in the evolution and  will represent, in the following, the response of the discrete surface to the presence of the walker.

\section{Numerical simulations}\label{sec:simulations}

\begin{figure}[ht!]
  \centering
  \includegraphics[width=\linewidth]{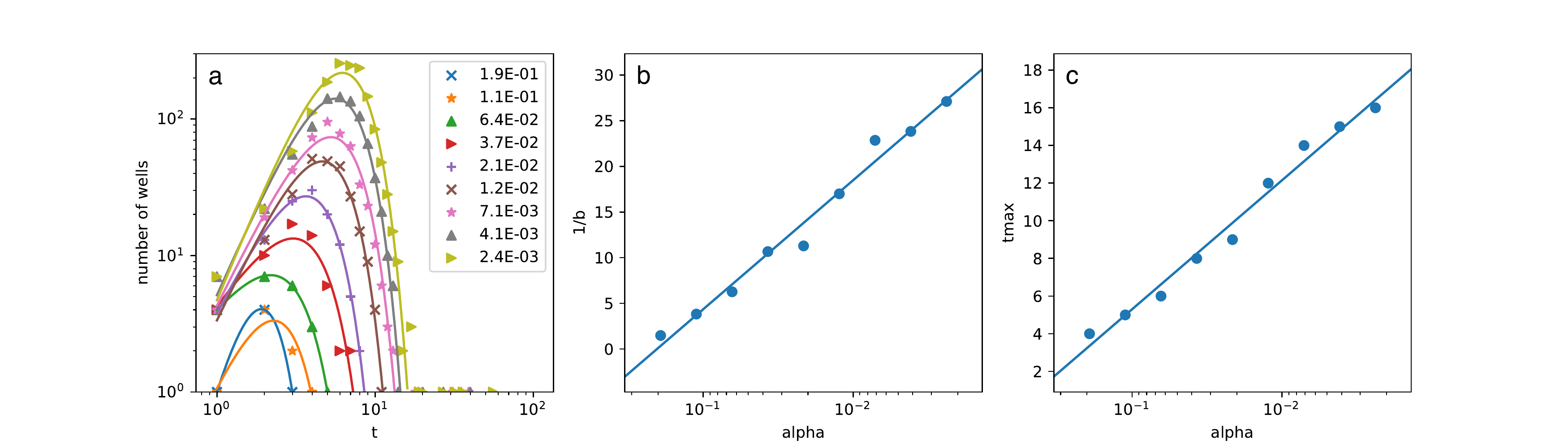}
   \includegraphics[width=\linewidth]{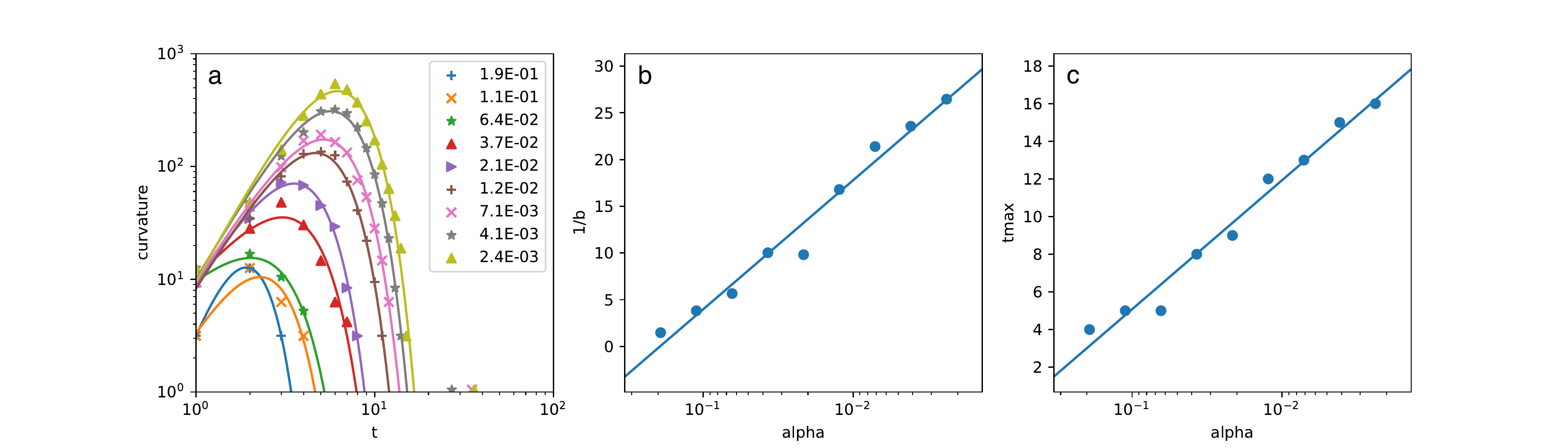}
  \caption{From the left to the right: (a) Number of wells (top) and  local curvature (bottom) in a ball of a radius 1, with $\beta = 3\alpha$, fit as $t^ae^{-bt^2}c$ ; (b) The parameter $b$ as function of $\alpha$; (c) The time $\texttt{tmax}$ of the exponential cut as function of $\alpha$.The initial condition $\psi(t = 0) = \frac{1}{\sqrt{3}}$ on each of the three internal components of the origin triangle and $0$ elsewhere.}
  \label{fig:wells}
\end{figure}

Although we have not derived the dynamical equations of the graph in the previous section, we can numerically characterize the growth of triangles, or more specifically the number of wells (in other words the number of $1\mathrm{-to-}3$ Pachner moves carried out minus the number of $3\mathrm{-to-}1$ Pachner moves carried out) in a region of finite space, centered around the initial condition. The reason to do that is to understand how and to what extent the metric responds to the walker when its probability density is more important. In fact, in the long run, since the probability density is lower, the response of the metric will be weaker and weaker until it disappears, which corresponds to an absence of Pachner moves.  For this reason we focus our study on a unitary radius ball. As depicted in Fig. \ref{fig:wells}.a, the number of wells behaves in the same way, whatever the values of $\alpha$ and $\beta$ chosen: \emph{(i)} first, it increases steadily, as in $t^a$, with $a > 0$ : the particle is localized and the surface stretches ; \emph{(ii)} second, it decreases very fast, as in $e^{-bt^2}$, $b > 0$ : the particle is not localized anymore and the surface relaxes ; \emph{(iii)} third, some small wells appear but disappear almost immediately ; \emph{(iv)} fourth, the surface is completely flat : the particle's localization is completely spread out leaving no place for Pachner moves. Notice that the exponential decrease ($e^{-bt^2}$) can be found in multiple models throughout complex systems and the constant $b$ is usually associated to an internal cut-off of the system. We thus suspected it was a function of $\alpha$. And indeed, plotting the value of $b$ against $\alpha$, as in Fig.\ref{fig:wells}.b, we can see that the constant $1/b$ depends linearly on the logarithm of $\alpha$ (where the ratio depends on the relationship between $\alpha$ and $\beta$), and similarly for the time at which the number of wells goes from decreasing exponentially fast to oscillating at small values, $\texttt{tmax}$. \\We will also see how curvature changes within this ball and not surprisingly, if we look at a vertex's curvature summed in the same ball of unitary radius, we recover exactly the same behavior as shown in Fig. \ref{fig:wells}. This is because the number of wells is strictly related to the local curvature. 
In particular, here we use a simple definition of the curvature concentrated in the deficit angles at the vertices, which is reminiscent of Regge calculus \cite{regge1961general, ambjorn1997geometry} : if a vertex of a graph is shared by $n$ triangles, then the vertex's curvature is equal to $2\pi - n\pi/3$. Notice that the global curvature is then the sum of the vertex's curvatures over the whole surface. For a 2-dimensional surface, it must be constant and vanishing, as stated by the Gauss-Bonet theorem. For instance in a triangular grid, where each vertex is shared by six triangles, the curvature is equal to $0$. Performing a $1\mathrm{-to-}3$ Pachner move on a triangle means adding negative local curvature $(-\pi)$ to the created vertex and positive local curvature $(\pi/3)$ to each of the three vertices of the former triangle, thus leaving the global curvature unchanged. In particular we consider that at time $t = 0$ the surface is globally and locally flat as in \cite{ArrighiTriang}, thus at all times the global curvature is equal to $0$.\\
Moreover, in all our simulations we choose $\beta = 3 \alpha$. In fact, we can distinguish two ranges : \emph{(i)} $\beta > 6\alpha$, that we call unstable and \emph{(ii)} $\beta < \alpha$, that we call quasi-stable. If $\beta > 6\alpha$, each $3\mathrm{-cycle}$ in the lattice is unstable, specifically the surface is always subject to a Pachner move at each time step. Indeed, if the probability of being inside the $3\mathrm{-cycle}$ is high enough (higher than $\beta$) so that we do not apply a $3\mathrm{-to-}1$ Pachner moves onto it, then there is a component $(v,k)$ inside it, with probability $\abs{\psi(t,v:k)}^2 > \beta / 6 > \alpha$, and consequently a $1\mathrm{-to-}3$ Pachner move is applied to $v$. If $\beta < \alpha$, no edge can be the exclusive cause of a $1\mathrm{-to-}3$ Pachner move once it was subject to a $3\mathrm{-to-}1$ Pachner move, by a similar reasoning. Our choice is therefore halfway between full instability and quasi-stability.

\begin{figure}[ht!]
  \centering
  \includegraphics[width=\linewidth]{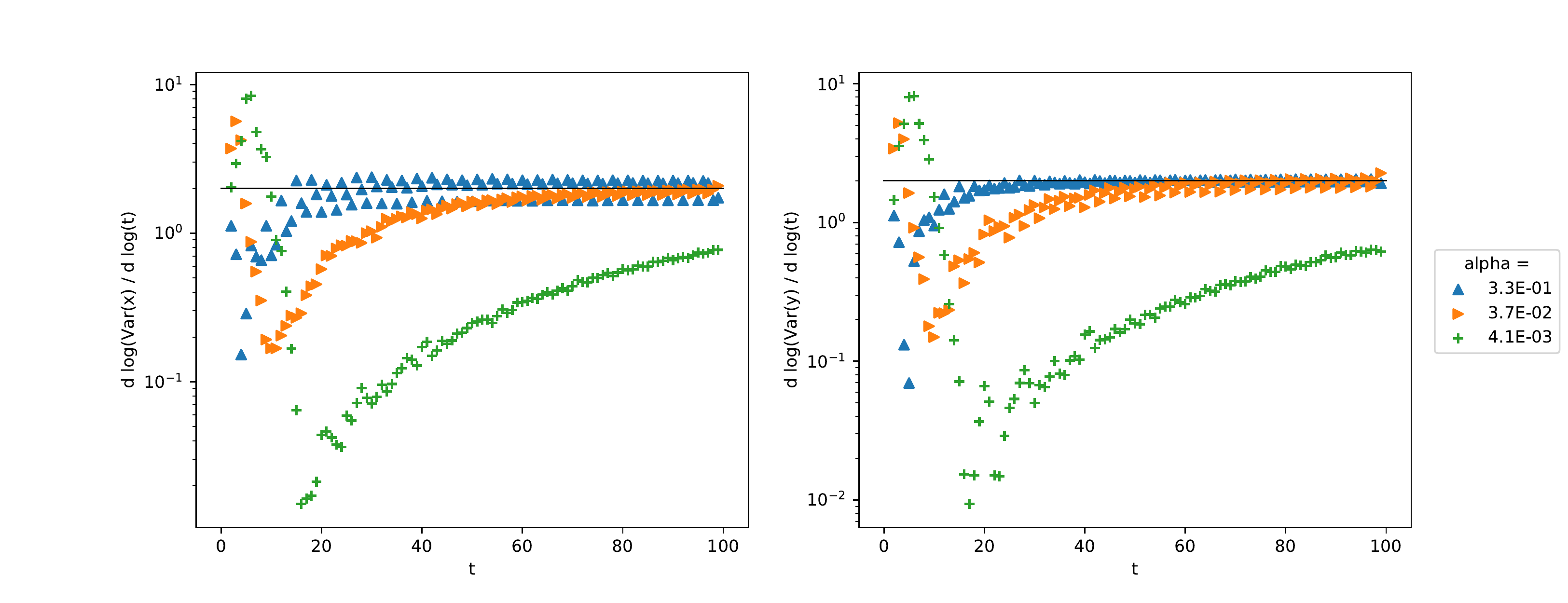}
  \caption{Gradient of the logarithm of the variance with $\beta = 3\alpha$ : it always converges to $2$. The initial condition $\psi(t = 0) = \frac{1}{\sqrt{3}}$ on each of the three internal components of the origin triangle and $0$ elsewhere.}
  \label{fig:variance}
\end{figure}

\begin{figure}[ht!]
  \centering
  \includegraphics[width=\linewidth]{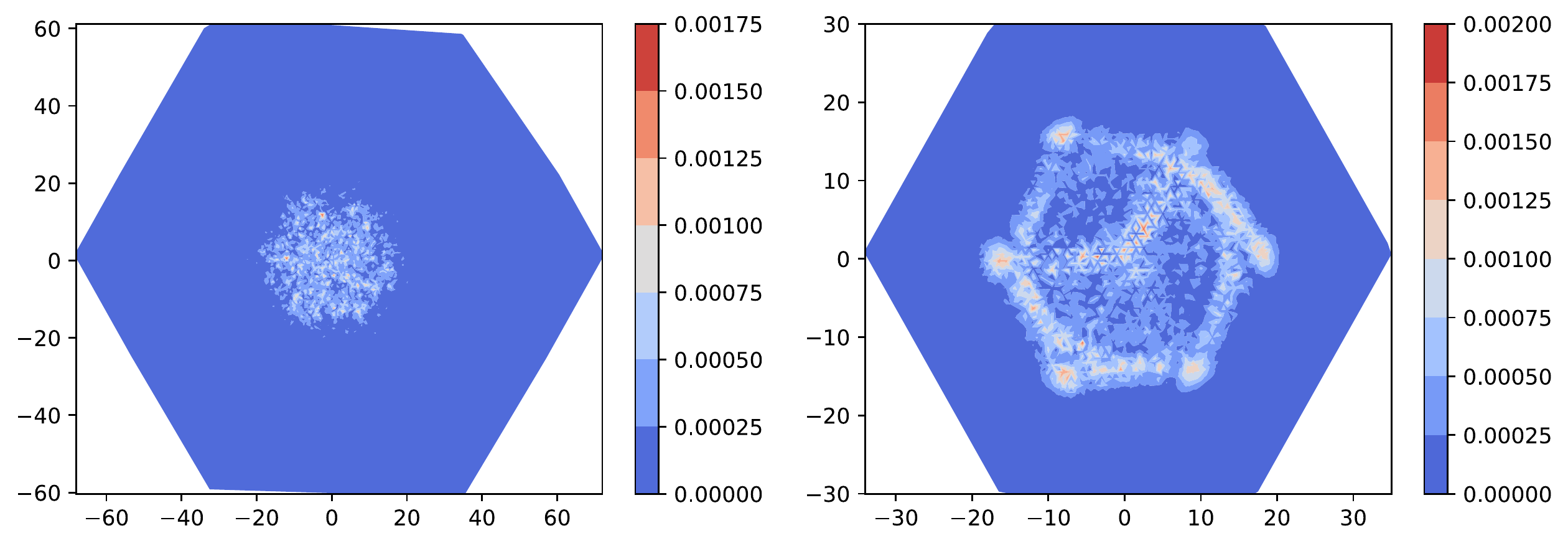}
  \caption{Heatmap after 100 steps of evolution. On the left $\alpha=10^{-4}$, on the right $\alpha=10^{-1}$. In both case $\beta=3\alpha$.\label{fig:longtermevolution}}
\end{figure}

Let us now concentrate on the walker dynamics. The impossibility to find an analytical solution of the equations \ref{eq: EQFIN}, does not prevent us to investigate numerically its dynamical properties. We have chosen to study its variance. If there are no Pachner moves, and therefore the metric is flat, it is well known that the variance is proportional to the square of the number of temporal steps, or otherwise said, the dynamics is ballistic. Thus, we measured the variance of its position $\Var[e] = E[e^2] - E[e]^2$, with $E[e] = \sum_e e |\Psi|^2$ the expectation value of the position of the walker along edge $e$. More precisely, because we expected that the variance be proportional to $t^\eta$, we chose to compute $\frac{\mathrm{d}\log{(\Var[e])}}{\mathrm{d}\log t}$, to have access to the exponent $\eta$. Notice that if the walk is symmetric with respect to $x$ and $y$, $\eta_x = \eta_y$. Because this condition holds for our walk, we will use $\eta$ as the global variance of the walker. In Fig.\ref{fig:variance} we can distinguish three regimes: long-, intermediate- and short-run regime. The long-run regime is not surprising: as soon as there are no Pachner moves the surface is completely flat and $\eta$ converges to $2$. In this regime, as one can also see in Fig. \ref{fig:longtermevolution}.b, the QW propagates as if there were no Pachner moves, and as one would expect on a regular triangular surface. In the short- and the intermediate-range regime the dynamics is more complex due to the non-linear coupling with the dynamical surface. In the first regime we can clearly see that for low values of $\beta$ and $\alpha$, the walker propagates hyper-ballistically. This is due to the fact that \threetoone Pachner move changes values of $\widetilde{\psi}$ arbitrarily far away from the edges that will be deleted. The lower the value of $\alpha$, the lower the probability required to realize a \onetothree Pachner move will be, which in turns implies that a greater number of \threetoone Pachner moves will happen. This can be summarized as follows: the lower the value of $\alpha$, the greater the response of the metrics will be. This also means that for lower values of $\alpha$, there will be a higher number of wells as we can see in \ref{fig:wells}.a, which results in a low variance in the intermediate-run regime, due to the fact that the walker takes much longer to get out of a finite region of space with a high number of wells. Note that in this intermediate-run regime the walker diffuses as one can see in Fig.\ref{fig:variance} and more specifically in Fig. \ref{fig:longtermevolution}.a.

\section{Summary and future work}

We introduced, to the best of our knowledge, the first example of a QW over a dynamical triangular grid, where rules changing the geometry are constructed using Pachner moves. 
Although, for simplicity, we restricted ourselves to $1\mathrm{-to-}3$ and $3\mathrm{-to-}1$ Pachner moves, we do believe it should be possible to extend our walker to $2\mathrm{-to-}2$ Pachner moves. We then wrote the equation that governs the behavior of the walker. It is non-linear, and represents the strong coupling between the dynamics of the walker and the grid : the walker evolves according to the state of the grid at time $t$, and the grid transforms according to the walker's probability density at the same instant. The wave function does not directly depend on the coordinates of the grid, which makes it prohibitive to compute the continuous limit. Still we think that a differential form can be derived in a mean field approximation or via coarse-graining maps, and we leave it for future research. 

Finally we found a robust local characterization of the curvature and the global number of wells, which has never been studied before, to the best of our knowledge. The evolution $t^a e^{-b t^2}$ is made of two factors : \emph{(i)} one is a power law dominant for small $t$, which accounts to the growing of the local curvature,  \emph{(ii)} the other is the exponential decay which eventually overwhelms the power-law behavior at very large $t$. Although it does not scale as a power law, it is a good approximation and in particular it naturally capture finite size effects of metrics. 
In future works, we would like to extend our model in multiple directions: \emph{(i)} considering multi-particle QCA, \emph{(ii)} consider higher dimensional simplicies, \textit{e.g.} tetrahedra, where global curvature is no longer constant, \emph{(iii)} consider a different way to make the $3\mathrm{-to-}1$ Pachner moves that would preserve the rotation symmetry as well as reversibility.

Finally, among the various extensions we hope for, there is the quantization of the underlying graph and its interaction with the QW. In fact, in our case the graph is purely classical. Similar models, by which we hope to be inspired for future research, are for example those considered in the Holstein polaron theory\cite{cruzeiro1994localized}, where there is a non-linear interaction between the electron and the quantum lattice which leads to deformation of the lattice. But even before we fully quantise our theory, we will have to make it reversible. In fact, as already said in the main part of the manuscript, here a $3\mathrm{-to-}1$ Pachner move is not the inverse operation of a $1\mathrm{-to-}3$ Pachner move, just because of the dependence on the probability of the walker's presence. In other words, even considering a quantum lattice, $1\mathrm{-to-}3$ Pachner moves and $3\mathrm{-to-}1$ Pachner moves cannot be seen as the analogous of a creation operator and a destruction operator. There is still a long way to go.

\authorcontributions{conceptualization, Q.A. and G.D.M.; methodology, Q.A.; software, Q.A.; validation, Q.A., N.E. and G.D.M.; formal analysis, Q.A. ; N.E. and G.D.M.; writing--original draft preparation, Q.A. and GDM; writing--review and editing, N.E. and G.D.M.; visualization, Q.A. and N.E.; supervision, G.D.M.; project administration, G.D.M.}

\funding{This work has been funded by the Pépinière d’Excellence 2018, AMIDEX fondation, project DiTiQuS.}

\acknowledgments{The authors acknowledge inspiring conversations with Pablo Arrighi and Nicolas Durbec and useful remarks on how to better present this work by the anonymous referees.}

\conflictsofinterest{The authors declare no conflict of interest.} 

\reftitle{References}
\externalbibliography{yes}
\bibliography{biblio}

\end{document}